# Simulating Venus' Cloud Level Dynamics Using a Middle Atmosphere General Circulation Model

Parish, H.F.[1] and J.L. Mitchell[1,2]

[1]*Department of Earth, Planetary and Space Sciences, University of California, Los Angeles, Los Angeles, CA 90095, USA*

[2]*Department of Atmospheric and Oceanic Sciences, University of California, Los Angeles, Los Angeles, CA 90095, USA*

## Abstract

The atmosphere of Venus is characterized by strong superrotation, in which the wind velocities at cloud heights are around 60 times faster than the surface rotation rate. The reasons for this strong superrotation are still not well understood. Motions in the atmosphere below the thick cloud deck are hard to determine remotely and in-situ measurements of the circulation below 40 km altitude are scarce. No model to date has been able to simulate superrotating winds with magnitudes comparable with those measured by entry probes, in the dense atmosphere between the surface and the clouds. However, important information on the dynamics and circulation of Venus' atmosphere can be determined by studying the atmosphere at cloud levels, where there are significantly more measurements than in the sub-cloud region, including the many recent observations made during the Venus Express and Akatsuki missions. In this work we describe a new Venus Middle atmosphere general circulation Model (VMM) to study the dynamics of the atmosphere at cloud altitudes. The model simulates the atmosphere from just below cloud deck to around 95 km altitude. We present simulations using the VMM with a simplified Newtonian cooling radiation scheme. Sensitivity studies have been performed to determine the most appropriate values for model parameters and the model has been validated by comparison with observations, including those from Venus Express and Akatsuki. The validated model provides some constraints on parameters which are poorly measured close to the boundary such as the mean winds and temperatures, and provides a basis for further investigations of the dynamics of Venus' cloud-level atmosphere. In future studies we will also investigate the influence of frequently-observed atmospheric waves, such as Kelvin and Rossby waves, to determine the role they play in generating the poorly-understood cloud-level structure at all latitudes.

## Background and Motivation

In this work, we describe a new Venus Middle atmosphere Model (VMM). The model has a lower boundary raised in height relative to the surface to just below cloud altitudes. The boundaries can be adjusted and in the simulations described here the lower boundary is at ~40 km altitude and the upper boundary is at ~95 to 100 km altitude. The approach of using an elevated lower boundary has been used successfully applied in many terrestrial models (Parish et al., 1992; Roble and Ridley, 1994; Walterscheid et al., 2000). We take this approach of developing a Venus middle atmosphere model for several important reasons. In-situ observations are scarce in Venus' lower atmosphere between the surface and the clouds. The only wind profiles in the 0 to ~40 km altitude range come from entry probes on the Pioneer Venus (Counselman et al., 1980) and Venera (Marov et al., 1973) missions. However, these profiles are limited in spatial and temporal coverage. They were all in the midnight to noon time frame and all but one was at latitudes between +/- 30◦ (Schubert, 1983). Measurements from the Visible and Infrared Thermal Imaging Spectrometer (VIRTIS) and Venus Radio Science Experiment (VeRa) instruments aboard Venus Express can be used to infer temperatures and winds within the ~40 to 90 km altitude range (Sanchez-Lavega et al., 2008; Piccialli et al., 2008, 2012; Hueso et al., 2012). Temperature measurements are available in the 0 to 10 km altitude range and composition is available within limited altitude ranges between 0 and ~35 km from VIRTIS (Baines et al., 2006; Svedhem et al., 2007). However, details of the circulation and dynamics below 40 km altitude are scarce. Significantly, the circulation and dynamics in the lowest scale height (~ 16 km) above the surface of Venus is very poorly known since the winds are small compared with the accuracy of the few measurements. In our modeling approach we therefore focus on the atmosphere at cloud heights, where there are significantly more measurements than in the sub-cloud region, especially since the Venus Express and Akatsuki missions.

The dynamics and circulation of Venus' atmosphere are also not well explained using existing models. No model to date has been able to generate superrotating winds in the dense atmosphere between the surface and the clouds with magnitudes comparable with those measured by the Pioneer Venus and Venera entry probes. Earlier general circulation models (GCMs) generated superrotation with magnitudes significantly smaller than those observed (Rossow, 1983; Del Genio et al., 1993; Del Genio and Zhou, 1996). Later GCMs generated cloud-level superrotation with magnitudes in the range of those observed using a simple thermal relaxation approach (Yamamoto and Takahashi, 2003a,b, 2004, 2006; Lee et al., 2005, 2007; Herrnstein and Dowling, 2007; Hollingsworth et al., 2007; Parish et al., 2011). However, in order to generate realistic winds, the heating rates used in the lower atmosphere needed to be significantly larger than those measured (Tomasko et al., 1980), and the superrotation was found to decrease dramatically if realistic heating rates were used (Hollingsworth et al., 2007; Yamamoto and Takahashi, 2009). More recently a full radiative transfer code with diurnal heating and realistic topography, included in a state-of-the-art Venus GCM (Lebonnois et al., 2010), generated cloud-level superrotation with wind magnitudes at the lower end of those observed. More recent simulations with enhancements to the physical processes show improvements to the superrotation below the cloud levels (Lebonnois et al., 2016). Other modeling efforts focussing on Venus' middle atmosphere have used different techniques, without an elevated lower boundary (e.g. Yamamoto and Takahashi, 2007, 2012). However, none of these models has produced superrotation between the surface and the clouds with magnitudes comparable to those observed by the Pioneer Venus and Venera probes. Lower atmosphere winds might be enhanced by adjusting the parameterization of surface-atmosphere interactions, but the values of relevant parameters have not been measured and there is no way to evaluate the realism of parameter choices. We therefore aim to realistically model the cloud-level atmosphere, without needing to accurately model the sub-cloud atmosphere.

Another motivation for developing a middle atmosphere model is the enormous gain in speed of generating results, since we obviate the necessity of performing simulations involving Venus' entire lower atmosphere between the surface and the cloud levels in order to generate realistic results at cloud heights and above. The large heat capacity of Venus' deep atmosphere makes simulations involving the entire lower atmosphere very computationally demanding. Even with current developments in computer processor speeds and massively parallel machines, simulations which involve taking the entire sub-cloud-level atmosphere of Venus to equilibrium can still take several weeks for a single simulation. By simulating only the cloud-level atmosphere and above, we can speed up our simulations by an order of magnitude. The extra speed with which we can perform simulations will allow us to thoroughly test sensitivities to atmospheric parameters and enable us to gain a much better understanding of Venus' cloud-level atmosphere and above than has been possible in the past, when we have been limited by the scarcity of measurements of the lower atmosphere and the necessity to simulate the entire lower atmosphere. Perhaps the only limitation of this approach is that we cannot self-consistently generate the zonal-mean circulation, since the lower atmosphere is not in place to maintain the circulation and input of momentum and energy from the lower atmosphere directly at the lower boundary of the middle atmosphere model. Thus, we will maintain realistic background values of winds and temperatures within the first ~2 km above the lower boundary using a simple linear friction approach. Above the frictional region, however, the atmosphere will be free-running, which will allow us to explore, for instance, the dynamic interactions of forced waves with the zonal-mean state.

There have been previous attempts to simulate Venus' atmosphere in the ~30 km to 110 km altitude range using an elevated lower boundary and wave forcing. For example, Newman and Leovy (1992) used a simplified spectral model and introduced tidal perturbations and Yamamoto and Tanaka (1997) forced 4 and 5 day waves at the lower boundary of a middle atmosphere model, making the simplifications of Newtonian cooling and Rayleigh friction. Yamamoto and Takahashi (2012) used a model in which they only time integrate the temperatures and winds above 30 km altitude and force 4 and 5.5 day planetary waves at the surface, and also use Newtonian cooling and Rayleigh friction. Imamura (2006) used a model with a lower boundary at 60 km altitude and forced different waves relative to fixed background winds and temperatures, rather than a freely circulating atmosphere, and used a Newtonian cooling approximation. In our simulations we use a Newtonian cooling simplification, in combination with free running background winds and temperatures above the region of forced lower boundary winds. Sensitivity tests will be performed to investigate the influence of different types of waves on the cloud-level atmosphere and results of simulations will be compared with observations.

## The Model

We have developed a new Venus middle atmosphere general circulation model based on the spectral version of the terrestrial Flexible Modeling System (FMS; Gordon and Stern, 1982). Using the model, we aim to simulate the dynamics of Venus' poorly-understood cloud-level atmosphere. The model makes use of a standard Eulerian spherical harmonic dynamical core, solving the nonlinear, global, time-dependent, hydrostatic primitive equations for an ideal gas over a sphere. The vertical coordinates are defined on a hybrid sigma-pressure Arakawa B-grid, with vertical differencing following Simmons and Burridge (1981). Prognostic variables include the zonal and meridional wind components, surface pressure, temperature, and vorticity and divergence of horizontal flow. We have adapted the model for Venus' atmosphere, by including appropriate physical constants, including the rotation rate, radius and gravitational acceleration for Venus. To focus attention on dynamical calculations within the Venusian cloud-level atmosphere, the lower boundary has been raised in height relative to the surface, to just below cloud altitudes. The approach of using an elevated lower boundary has been used successfully in many terrestrial models (e.g. Parish et al., 1992; Roble and Ridley, 1994; Walterscheid et al., 2000). The upper and lower boundary heights can be adjusted and varied as required.

In the simulations described here, the lower boundary height has been set to around 40 km altitude ($\sim 4$ x $10^5$ Pa), which is below the cloud deck at an altitude low enough that conditions close to the boundary do not directly impact the region of interest at cloud heights, but within an altitude range where there are sufficient observations to make reasonable estimates of parameters such as winds and temperatures at the lower boundary. The lower boundary is also in a comparatively stable region; the atmosphere between around 35 km to 45 km altitude has been found from Pioneer Venus (PV) and Venera observations to have relatively high static stability (Schubert, 1983; Gierasch et al., 1997), below more turbulent regions just below the cloud deck. Observations within the ~40 to 45 km altitude range have been used to estimate atmospheric temperatures and winds in this altitude range. These included profiles of winds and other atmospheric parameters from the PV and Venera probes (Schubert et al., 1980; Counselman et al., 1980; Gierasch et al., 1997) for limited ranges of latitude and local times, radio occultation measurements from PV and from the VeRa radio science experiment on Venus Express, which provide temperature measurements down to around 40 km altitude (Seiff et al., 1980; Tellmann et al., 2009, 2012). Wind values in the same altitude range inferred from these measurements via the thermal wind equation, assuming cyclostrophic balance and measurements from the Venus Express VIRTIS instrument also provide winds down to around 40 km altitude (Hueso et al., 2012). The upper boundary is currently around 95 to 100 km altitude (~3 Pa). A simple linear friction was introduced within the first ~2 km from the lower boundary, as discussed below. A model schematic is shown in Figure 1. The radiation scheme we use here is a simplified Newtonian cooling radiation approximation, based on the Held-Suarez (HS) formulation, for relaxation of temperature to a specified radiative equilibrium (Held and Suarez, 1994). The radiative cooling timescale is set to 40 days above the first 2 km, which is similar in magnitude to estimated timescales in the upper atmosphere of Venus derived from observations (e.g. Crisp, 1986). The horizontal resolution in the model can be varied. In the simulations described here a T21L31 resolution has been used (64 longitudes and 32 latitudes), with 31 sigma-pressure vertical levels.

## Baseline Simulations

Baseline simulations were performed, with a relaxation temperature profile based on observations (e.g. Schubert et al., 1980; Tellmann et al., 2009; Imamura et al., 2017). Winds were initialized at rest at all altitudes and simulations were run to equilibrium for $\sim$ 2100 Earth days (~18 Venus days). Within the Held-Suarez approximation, the tropopause temperature ($T_t$) was initially set to 210 K, based on the range of observational values (e.g. Schubert et al, 1980) and the parameter that controls the variation of potential temperature with height ($\Delta\theta$) was initially set to 25 K (Run 1). A list of runs and their corresponding parameter values is given in Table 1. The results of simulations are shown in Figures 2a to 2c.

### 1) Temperature

Figure 2a shows contours of the simulated zonally-averaged temperature as a function of latitude, pressure level and approximate height. Temperatures range from around 420 K at ~40 km altitude to around 210 K at ~75 km altitude, consistent with observations (Schubert el al., 1980; Tellmann et al, 2009, Imamura et al., 2017). Latitudinal temperature contrasts are small, around 5K, at the lower end of the range of Pioneer Venus and Venera probe observations, which show temperature differences of around 5 to 10 K up to ~55 km altitude, and smaller than the measured variations of around 15 to 20 K at higher altitudes (e.g. Schubert et al., 1980).

### 2) Zonal Winds

The simulated zonally-averaged zonal wind (positive westward) is shown in Figure 2b for the same simulation as a function of latitude, pressure and approximate height. The zonal wind displays westward jets at high latitudes around $70^0$ to $75^0$ latitude, and at altitudes between ~60 to 65 km, with smaller magnitudes at lower latitudes, and a smaller eastward circulation at the lowest levels. The height of the maximum wind is lower than that observed on Venus at ~65 to 70 km altitude (e.g. Schubert et al., 1980). The simulated wind magnitudes are significantly too small, maximizing at less than ~10 m/s, compared with the very strong winds measured using the Pioneer Venus and Venera probes between 50 to 100 m/s at 65 km altitude (Schubert et al., 1980) and cloud-tracked winds of over 100 m/s (Schubert et al., 1980; Del Genio and Rossow, 1990). The observed zonal winds also tend to maximize at the equator and low latitudes (e.g. Sanchez-Lavega et al., 2008) rather than at high latitudes.

### 3) Meridional winds

The corresponding meridional winds (positive northward) are shown in Figure 2c. The meridional wind is directed poleward from the equator in each hemisphere at the altitude of the zonal jets, consistent with Pioneer Venus and Mariner 10 observations (Suomi 1974, Limaye and Suomi, 1981). Simulated meridional wind magnitudes are up to around 3 m/s at this altitude, a little smaller than observed in Pioneer Venus measurements in which they are in the range of ~5 to 10 m/s (Rossow et al., 1980). At lower altitudes, simulated meridional wind magnitudes are around 1-2 m/s, compared with magnitudes of a few m/s in observations (Counselman et al.,1980). A series of layered, meridional cells are present below around 55 km altitude.

## Lower Boundary Drag

The winds at the lower boundary of the model do not have magnitudes comparable with measured values, since the lower atmosphere is not simulated in this model and the angular momentum of the lower atmosphere is not available to keep lower boundary wind magnitudes consistent with those observed. A simple linear friction, with frictional time constant set to 1 day, was introduced within the first ~2 km from the lower boundary to maintain zonal and meridional winds within the range of measured values at these altitudes. Realistic zonal and meridional winds were estimated using values observed from Pioneer Venus, Venera and Venus Express (Schubert et al., 1980; Sanchez-Lavega et al., 2008; Hueso et al., 2015). Above the narrow region of lower boundary friction the atmosphere is free running and atmospheric variables vary according to the general circulation of the model. Winds were initially set at rest at all altitudes above the region of lower boundary drag. Simulations including this lower boundary drag were run to equilibrium for around 19 Earth years (~60 Venus days). In this example (Run 2) the tropopause temperature, $T_t$, was chosen to be 210 K and the change in potential temperature with height parameter ($\Delta\theta$) was set to 25 K, where the parameters we have used are listed in Table 1. The effects of introducing this lower boundary drag are discussed below.

*a) Temperature*

Figure 3a shows the simulated zonally-averaged temperature as a function of latitude, pressure and approximate height, with lower boundary drag included. The range of simulated temperatures between ~ 200 and 420 K is consistent with Pioneer Venus and Venera probe observations (Schubert et al., 1980). Latitudinal temperature contrasts below ~55 km altitude are ~5K, and ~20K at ~60 km altitude, which is comparable with Pioneer Venus and Venera measured temperature contrasts within these altitude ranges (Schubert et al, 1980). There are temperature inversions at altitudes above ~100 mbar in the region of the tropopause, which occur at higher pressures at high latitudes, consistent with Pioneer Venus measurements (Schubert et al., 1980).

*b) Zonal Winds*

The corresponding zonally-averaged zonal winds (positive westward) are shown in Figure 3b. There is a broad maximum in the zonal winds which runs across the equator at a height of ~65 to 70 km altitude, at a similar altitude to that observed by the Pioneer Venus and Venera probes (Schubert et al., 1980), by Venus Express (Sanchez-Lavega et al., 2008; Hueso et al., 2015) and in Akatsuki measurements (Horinouchi et al., 2017). Zonal wind magnitudes become smaller at higher latitudes. The largest magnitudes are over 100 m/s, comparable with those found in the probe, Venus Express and Akatsuki measurements.

*c) Meridional Winds*

Meridional winds are shown in Figure 3c. The simulated meridional winds are small in magnitude, generally less than ~ 3 m/s, which is comparable with observed magnitudes of a few m/s observed by Pioneer Venus probes below around 50 km altitude (Counselman et al.,1980) and tend to show a layered meridional cell structure below around 50 km altitude, which is consistent with Pioneer Venus probe observations (Schubert et al.,1980) which suggest alternating meridional wind directions with altitude in this height range. Simulated magnitudes are a little smaller than magnitudes observed by Pioneer Venus above that altitude, which are in the range of ~5 to 10 m/s (Rossow et al., 1980), although meridional winds in the 47 to 66 km altitude range are variable in Venus Express measurements (Hueso et al., 2015) and have estimated errors of around +/- 5 to 10 m/s, which is comparable in magnitude to the meridional winds themselves.

## Comparison with Observations

The vertical and latitudinal structure of the same simulations including lower boundary drag are compared with observations below.

### a) Effects on Temperature

*(i) Vertical Profile*

The simulated vertical temperature profiles at $10^0$S and $70^0$N with and without lower boundary drag are compared with observations at the same latitudes, in Figures 4a and 4b respectively. Figure 4a shows the simulated temperature profile at $10^0$S compared with Venus Express (Tellmann et al., 2009), Venera 8 (Schubert et al., 1980) and Akatsuki observations (Imamura et al., 2017). The simulations with and without lower boundary drag both show reasonable agreement with the observed profiles within the variability of the different measurements, although they are too low around 100 mbar by up to ~20K between ~60 km to 70 km altitude, compared with the Venus Express and Akatsuki measurements. Given the simplicity of the radiative approximation, the agreement is relatively good at most altitudes.

The simulated temperature profiles at $70^0$N are compared with Venus Express (Tellmann et al, 2009) and Magellan (Jenkins et al., 1994) observations in Figure 4b. The simulated temperatures show reasonable agreement with the observations to within around 20K at most altitudes. The simulations without lower boundary drag show closer agreement with the Magellan observations. However, the simulations with lower boundary drag are closer to the Venus Express measurements above around 65 km and also show some similarities in the temperature minimum observed between around 60 to 65 km altitude in the Venus Express measurements, which may be due to the cold collar often observed at this altitude and latitude (Taylor et al., 1979, 1980; Piccioni et al., 2007). The presence of a cold collar will be explored in future simulation, but is beyond the scope of this work.

*(ii) Latitudinal Profile*

The latitudinal variation of the simulated temperature, with and without lower boundary drag, is compared with Pioneer Venus measurements (Schubert et al., 1980) from the equator to the pole at ~60 km altitude (Figure 4c). At this altitude, there is little variation of the temperature with latitude without lower boundary winds. With lower boundary winds included, simulated temperatures show a decrease towards higher latitudes of about 20K. The measurements are around 20K higher at this altitude than the simulated values and also show an overall decrease of around 20K from equator to pole, but with a large drop in temperature of around 40K in the range of $50^0$ to $80^0$ latitude, likely associated with the high latitude cold collar region (Taylor et al., 1979, 1980; Piccioni et al., 2007). The complexity of the high latitude regions of Venus' cloud-level atmosphere are not well-reproduced in our current idealized representation. The complex structure at high latitudes is not well understood and may be related to the presence of waves or other phenomena. In future work, we plan to include the effects of various waves, such as Kelvin and Rossby waves, which have regularly been seen in Venus' cloud-level atmosphere (e.g. Del Genio and Rossow, 1990) and may lead to more complex latitudinal structure (e.g. Ando et al., 2017; Kashimura et al., 2019), although this is outside the scope of this work. The magnitude of the temperature and sense of its variation away from the equator is generally improved by the presence of lower boundary drag.

## b) Effects on Zonal Winds

### (i) Vertical Profile

The corresponding simulated vertical profiles of zonal wind velocity are compared with observations at $5^0$ and $30^0$ latitude in Figures 4d and 4e respectively. In Figures 4d and 4e simulated wind profiles are compared with winds measured by Pioneer Venus (Schubert et al., 1980), and by Venus Express in ultraviolet, visible and near infrared, at ~66 km, 61 km and 47 km altitude respectively (Sanchez-Lavega et al, 2008; Hueso et al., 2015), and by Akatsuki (Horinouchi et al., 2017).

The vertical profiles of the simulated zonal winds show magnitudes at both $5^0$ and $30^0$ latitude which fall within the general variability of the Pioneer Venus, Venus Express and Akatsuki observations. In comparison, the simulated zonal winds without lower boundary drag are very much smaller at both $5^0$ and $30^0$ latitude, and are around 1-2 m/s. The magnitudes of the simulated zonal winds where lower boundary drag is included, tend to be underestimated around 40 to 45 km altitude and overestimated around 50 to 60 km altitude, although errors on the Venus Express zonal wind measurements in the 47 to 66 km altitude range are estimated to be in the range of around +/- 15 to 20 m/s (Hueso et al., 2015).

### (ii) Latitudinal Profile

The simulated variations of the zonal winds with latitude at ~ 66 km altitude are compared with Venus Express observations in Figure 4f. Averaged Venus Express measurements at ~66 km altitude are shown for periods between 2006 and 2008 and between 2009 and 2011 respectively and range from around 100 to 120 m/s at latitudes < $50^0$ to ~20 to 30 m/s at $80^0$ latitude. The average Venus Express winds showed a long term increase between 2006 to 2008 and 2009 to 2011 at latitudes < ~ $50^0$ (Khatuntsev et al., 2013; Hueso et al., 2015). (Note, however, the significant change in the wind velocity from 2006 to 2010 reported by Khatuntsev et al. (2013) is now believed to be an aliasing artifact; see Bertaux et al., 2016). Simulated winds with lower boundary drag included show magnitudes in between those in the earlier and later intervals at latitudes < $50^0$. Zonal winds at this altitude without lower boundary drag included are very much smaller than the measured winds (~5 m/s or less). Zonal winds with lower boundary drag included are overestimated between ~$50^0$ and $80^0$ latitude. This may be related to the influence of waves or other processes which are not represented in this simplified model, and this will be the subject of future investigations.

## Sensitivity Experiments

Sensitivity experiments were performed to determine the influence of changes in model parameters on the dynamics at cloud altitudes. We discuss the sensitivity to changes in the lower boundary drag wind magnitudes, and the effects of changing the tropopause temperature and the change of potential temperature with height. We compare the results of simulations with observations.

### 1) Sensitivity to Changes in Lower Boundary Winds

The lower boundary drag is set to maintain the zonal and meridional winds with values consistent with measurements around the altitude of the lower boundary (~40 km altitude) (e.g. Schubert et al., 1980; Sanchez-Lavega et al, 2008, Hueso et al., 2015), with frictional time constant 1 day. The sensitivity of the response to changes in the magnitudes of these winds was examined, by reducing the magnitudes of the lower boundary winds to half of their original values (Run 3 in Table 1), whilst keeping other parameters the same. Simulations were run to equilibrium for 19 Earth years (~60 Venus days), from an initial state with winds at rest at all altitudes above the region of lower boundary drag.

### a) Effects on Temperature

#### (i) Vertical Profile

The changes in the simulated vertical temperature profiles at $10^0$S and $70^0$N when the magnitudes of the lower boundary drag winds are reduced to half their original magnitudes are shown in Figures 5a and 5b, and compared with the same Venus Express, Akatsuki and Venera 8 measurements shown in Figures 4a and 4b. Adjusting the lower boundary winds has a relatively small effect on the temperature profile at $10^0$S or $70^0$N, changing the temperature by around 10 to 15K or less.

#### (ii) Latitudinal Profile

Figure 5c shows the latitudinal variations of the simulated winds for the adjusted magnitude of lower boundary drag winds, compared with the same Pioneer Venus observations shown in 4c. The simulated zonal winds with the original drag magnitude show a larger latitudinal temperature difference of around 30K from the equator to pole, compared with around 5 - 10K for the half magnitude drag, although in both cases the simulated latitudinal variations are significantly less pronounced than those found in the observations using this simplified approach. High latitude variations will be investigated in future model simulations.

**b) Effects on Zonal Winds**

*(i) Vertical Profile*

Figures 5d and 5e show the simulated vertical zonal wind profiles at $5^0$ and $30^0$ latitude respectively, for simulations with the original lower boundary drag wind decreased to half of their original magnitudes. The results of the simulations are compared with the same observations shown in Figure 4d and 4e from Pioneer Venus, Venus Express and Akatsuki. The zonal wind response at altitudes up to the zonal wind maximum around 65 km altitude is approximately proportional to the magnitude of the lower boundary winds, although the zonal wind magnitudes fall off less rapidly at altitudes above 65 km for the original lower boundary winds than for the winds of half magnitude, at both $5^0$ and $30^0$ latitude. The magnitudes of the simulated response are most consistent with the observations when the original lower boundary wind values (Run 2) are used.

*(ii) Latitudinal Profile*

The effects of changing lower boundary drag on the latitudinal variation of the simulated zonal winds are shown in Figure 5f. Comparison is made with the same Venus Express observations shown in Figure 4f. The simulated zonal wind magnitudes are decreased approximately in proportion to the magnitude of the lower boundary drag winds, generating zonal winds in Run 3 of around half the magnitude of those generated in Run 2, at all latitudes. The zonal winds simulated in Run 2 are much closer in magnitude to the observed winds than those simulated in Run 3, up to around $60^0$ latitude. As noted above, the zonal winds tend to be overestimated at high latitudes in our simplified approach.

**2) Effects of Changes in Tropopause Temperature**

**a) Effects on Temperature**

*(i) Vertical Profile*

The tropopause temperature and altitude shows significant variability and is latitude-dependent (e.g. Tellmann et al., 2009; Jenkins et al., 1994; Schubert, 1983). The tropopause temperature tends to decrease towards higher latitudes with a minimum centered around $70^0$ latitude (Tellmann et al., 2009). Since there is variability in the tropopause temperature, the sensitivity of the simulated cloud-level temperatures and winds to changes in the tropopause temperature was investigated in model simulations. In these simulations the lower boundary drag was the same as in Run 2 and all parameters other than the tropopause temperature were kept the same. A constant value of the change in potential temperature with height ($\Delta\theta$) of 10K was used. In each case simulations were run to equilibrium for 19 Earth years (~60 Venus days), from an initial state with winds at rest at all altitudes above the region of lower boundary drag.

Figures 6a to 6c show vertical and latitudinal temperature profiles of the simulated temperature, for three different values of tropopause temperature ($T_t$), i.e. 220K, 240K and 260K (Runs 4, 5, and 6 in Table 1 respectively). Simulated vertical temperature profiles are shown in Figures 6a and 6b at $10^0$ and $70^0$N respectively and are compared with the same occultation data shown in Figures 4a, 4b and 4c, from Magellan, Venus Express, Venera 8 and Akatsuki. For the simplified radiation approximation we are using in here, the simulated temperature profiles at $10^0$ and $70^0$ latitude are similar up to ~55 km altitude for all values of $T_t$, but differ above that altitude according to the chosen tropopause temperature, where the simulations with higher tropopause temperatures turn off to a constant temperature at lower altitudes. There are differences between the observations from different sources of between ~10 to 40 K at altitudes between 40 km and 75 km.

At $10^0$S (Figure 6a) the simulated temperatures for all $T_t$ are close to the Venera 8 observations up to around 50 km altitude. Between around 50 and 60 km altitude, simulated temperatures and are lower than observations by around 20 to 30K. The simulated profile has a different shape from that of the Venus Express and Akatsuki observations above around 55 km altitude. The $T_t = 260$K simulations are closest to the Venus Express and Akatsuki observations around 65 km altitude, but the $T_t = 220$K and 240K simulations are closest to the Venera 8 observations at the same altitude. The $T_t = 220$K simulations are closest to the observations at ~75 km altitude.

At $70^0$N (Figure 6b) simulated values for $T_t = 220$ K are with around 20K of the Venera and Magellan observations at all altitudes, which is comparable with the differences between the Venus Express and Magellan observations themselves. The Venus Express measurements show a slight minimum of ~10 to 20K just below ~65 km altitude, which is not seen in the Magellan observations nor the simulations. The simulations for $T_t = 240$K and 260K show temperatures which are too large by ~20 to 40K between 65 and 75 km altitude.

*(ii) Latitudinal Profile*

The latitudinal variations of temperature at ~65 km altitude compared with Pioneer Venus observations are shown in Figure 6c. The observations at this altitude show a temperature drop of around 20K between the equator and the pole, with a local minimum around $60^0$ to $70^0$ latitude, likely associated with the polar cold collar. Simulated temperatures show magnitudes within the range of observations, although the complexities of the latitudinal variations of temperature are not well captured with the simplified radiation scheme we use here. As noted above, the nature of Venus' high latitude circulation is currently not well understood, and we plan to investigate the high latitude region further in future work.

**b) Effects on Zonal Winds**

*(i) Vertical Profile*

Figures 6d and 6e show the effects of changes in tropopause temperature on the vertical profile of the zonal winds at $5^0$ and $30^0$ latitude respectively. Simulated winds are compared with the same observations as in Figures 4d and 4e. The effects of changes in tropopause temperature on simulated zonal winds are most significant above about 55 km altitude. Zonal winds above this altitude increase by ~10 to 20 m/s as the tropopause temperature is decreased from 260K to 240K and as it is decreased from 240K to 220K, at both $5^0$ and $30^0$ latitude. There are considerable differences between different zonal wind observations. The simulated zonal winds are generally within the range of variation of the measured values for all values of $T_t$ up to ~60 km altitude, but tend to be too low in magnitude at higher altitudes. As the tropopause temperature is increased smaller zonal wind magnitudes are produced above ~55 km altitude. The simulated zonal winds show evidence of a small local maximum at a similar height to that seen in the Pioneer Venus observations at $5^0$ latitude around 50 km altitude, for all values of $T_t$.

*(ii) Latitudinal Profile*

The latitudinal variations of the simulated zonal winds (Figure 6f) are compared with the same Venus Express observations used in Figure 4f, for the different values of $T_t$. For all magnitudes of tropopause temperature the zonal winds decrease from the equator to the pole by ~ 70 - 80 m/s. The observed zonal winds show similar equator to pole decreases. The overall zonal wind magnitude decreases as the tropopause temperature increases at all latitudes, and is closest in magnitude to the observed values for $T_t$ = 220K. The zonal wind variation with latitude is not as complex in the simulations as in the observations for the simple approach we use here; for example a small maximum is observed in the zonal winds around $40^0$ to $50^0$ latitude, which is not reproduced in the model simulations. The introduction of waves of different kinds or a more complex radiation scheme may improve the simulated structure in future simulations.

**3) Sensitivity to Potential Temperature Change with Height**

The sensitivity to variations in the change of potential temperature with height ($\Delta\theta$) were also investigated. Figures 7a to 7f show vertical and latitudinal temperature profiles of the simulated temperatures and zonal winds, for different values of the parameter that controls the change of potential temperature with height ($\Delta\theta$), i.e. for $\Delta\theta$ = 5K, 15K and 25K (Runs 7 to 9 in Table 1). In all cases the same lower boundary drag used in Run 2 was included and all other parameters were kept the same. In this case a value of $T_t$ = 220K was used. In each case simulations were run to equilibrium for 19 Earth years (~60 Venus days). Winds were initially at rest at all altitudes above the region of lower boundary drag.

**a) Effects on Temperature**

*(i) Vertical Profile*

Figures 7a and 7b show simulated vertical temperature profiles at $10^0$S and $70^0$N respectively for the three $\Delta\theta$ values of 5K, 15K and 25K. Simulations are compared with the same occultation data, as shown in Figures 4a and 4b from Magellan, Venus Express, Venera 8 and Akatsuki. At $10^0$S (Figure 7a) simulated temperatures are within ~10 - 20K of those observed, at all altitudes. The temperature at a given altitude increases as $\Delta\theta$ increases and is closest to the observed temperature at all altitudes for $\Delta\theta$ = 25K. At $70^0$N (Figure 7b) the simulated profiles are similar for all $\Delta\theta$ and are generally 10-20 K cooler than observations up to ~65 km latitude and are close in magnitude to the Venus Express observations between around 65 and 75 km altitude.

*(ii) Latitudinal Profile*

The simulated latitudinal temperature variations are compared with Pioneer Venus observations at ~65 km altitude in Figure 7c. Simulated temperatures vary by around 15 to 20K between the equator and the pole. Observations show a difference of around 20K between the equator and the pole, but show much stronger structure, with a large minimum centered around $65^0$ latitude, likely associated with the cold collar region, as noted above. The highest temperatures at the equator correspond to the highest $\Delta\theta$, and the opposite is true at higher latitudes. All of the simulations with this simplified radiation scheme show less latitudinal structure than observed, as also noted above.

**b) Effects on Zonal Winds**

*(i) Vertical Profile*

The vertical profiles of simulated zonal wind magnitudes for the same three values of $\Delta\theta$ (see Table 1) are shown in Figures 7d and 7e, at $5^0$ and $30^0$ latitude respectively. Observations from Pioneer Venus, Venus Express and Akatsuki are shown for comparison. In all cases, higher $\Delta\theta$ significantly enhances the magnitudes of the peak winds, and is closest to the observed winds in this case for $\Delta\theta = 15$K at both latitudes.

*(ii) Latitudinal Profile*

The simulated latitudinal variations of zonal wind are shown in Figure 7f, compared with the same Venus Express measurements as shown in Figure 4f. Significantly higher wind velocities are produced by large $\Delta\theta$, especially at lower latitudes. The zonal wind magnitudes for $\Delta\theta = 15$K are closest in magnitude to those observed, with an equator to pole difference in zonal wind magnitude of around 110 m/s. The simulated wind magnitudes are generally larger than those observed at high latitudes, and do not show the details of the latitudinal structure seen in the observations, e.g. the small zonal wind magnitude peak at around $40\text{-}50^0$ latitude. A more sophisticated radiation code and perhaps the presence of different types of waves as well as an understanding of the dynamics of the complex polar regions may be required to generate structure similar to that observed, and will be the subject of future work using the model.

## Table 1: Parameters used in Model Simulations

| Run Number | Tropopause temperature, $T_t$ (K) | Vertical change in potential temperature, $\Delta\theta$ (K) | Lower boundary drag |
|:---:|:---:|:---:|:---|
| 1 | 210 | 25 | None |
| 2 | 210 | 25 | Original |
| 3 | 210 | 25 | Original x 0.5 |
| 4 | 220 | 10 | Original |
| 5 | 240 | 10 | Original |
| 6 | 260 | 10 | Original |
| 7 | 220 | 5 | Original |
| 8 | 220 | 15 | Original |
| 9 | 220 | 25 | Original |
|  |  |  |  |

## Discussion and Conclusions

We have constructed a Venus Middle atmosphere Model (VMM), based on the well-known FMS spectral dynamical core, so we can focus on processes within the cloud-level atmosphere between around 40 km and 95-100 km altitude. The model includes a simplified Newtonian cooling radiative forcing, and realistic winds are introduced at the lower boundary of the model using a simple linear friction approach to maintain winds consistent with observations, in the absence of the momentum source of the lower atmosphere. Above the narrow region of lower boundary friction, which is maintained within the first ~2km, the atmosphere is free running and atmospheric variables vary according to the general circulation of the model.

We have performed sensitivity studies with VMM, to determine the effects of varying different parameters within the model. We find that winds and temperatures are sensitive to changes in the model parameters of tropopause temperature, changes in the potential temperature with height and the magnitudes of the lower boundary winds. Our sensitivity tests show that when we vary the lower boundary wind magnitudes, the effects on the temperature are relatively small, but more structure is seen in the variation of the temperature with height and latitude that show more similarity with observed variations than the case without lower boundary winds. Changing the magnitude of the lower boundary winds has significant effects on wind magnitudes in the free-running part of the atmosphere at all heights. The peak zonal wind magnitudes are approximately proportional to the magnitudes of the lower boundary winds.

We have also tested the sensitivity to changes in the tropopause temperature, $T_t$, which has some variability in the observed values (e.g. Tellmann et al., 2009). We experimented on tropopause temperatures within the broad range of observed values, because the simplified Newtonian type radiation in the model may not reproduce all of the features of the vertical and latitudinal variations in the tropopause temperature, but performing sensitivity tests within the model allows us to determine the effects of variations in this parameter. The effects of changes in the tropopause temperature on the simulated temperatures are mainly at altitudes above around 55 km, and primarily control the maximum temperature in the 65 km to 75 km altitude range. A higher tropopause temperature also tends to reduce the peak zonal wind magnitude significantly. If the tropopause temperature is larger, the tropopause occurs at lower altitudes and the equator-to-pole temperature gradient becomes relatively small at those altitudes. Thus the dynamical driving of winds due to the latitudinal temperature gradient is relatively small at those altitudes, and the magnitudes of the zonal winds are reduced.

We have also tested the sensitivity to changes in the parameter that modifies the potential temperature variation with height ($\Delta\theta$), which has an effect on the overall temperature lapse rate. For higher values of $\Delta\theta$, there is a slower drop of temperature with height, although the tropopause temperature remains the same. The changes in the decrease of temperature with height are mainly at lower latitudes and below around 65 km altitude. Peak zonal wind magnitudes are greatly increased for higher values of $\Delta\theta$. When there is a slower drop of temperature with height, occurring mainly at lower latitudes, a larger equator-to-pole temperature difference is produced, which drives stronger zonal winds.

We have shown that we are able to simulate a superrotating atmosphere with basic features which compare favourably with observations using a very simple model. The model is forced using a Newtonian radiation approximation and a simple lower boundary drag to maintain winds close to measured values within ~2 km of the lower boundary, with a free running atmosphere at higher altitudes. There are a small number of variable parameters within the model. The choice of parameters for $T_t$ and $\Delta\theta$ for Run 1, for example, gives a temperature structure and zonal winds magnitudes which are generally within the range of variability of the observations. Our sensitivity experiments indicate how sensitive the simulated results are to variations in those parameters. The validated VMM model is now available to provide a basis for future investigations. The simplified model does not reproduce all of the detailed features of the observations, for example the structure at high latitudes, which is still poorly understood. We may reasonably expect more complex features need to be added to the model, for example a more sophisticated radiative scheme, or different types of atmospheric waves, many of which are regularly observed in the cloud-level atmosphere. We plan to investigate these phenomena using VMM in future work.

# **Figures**

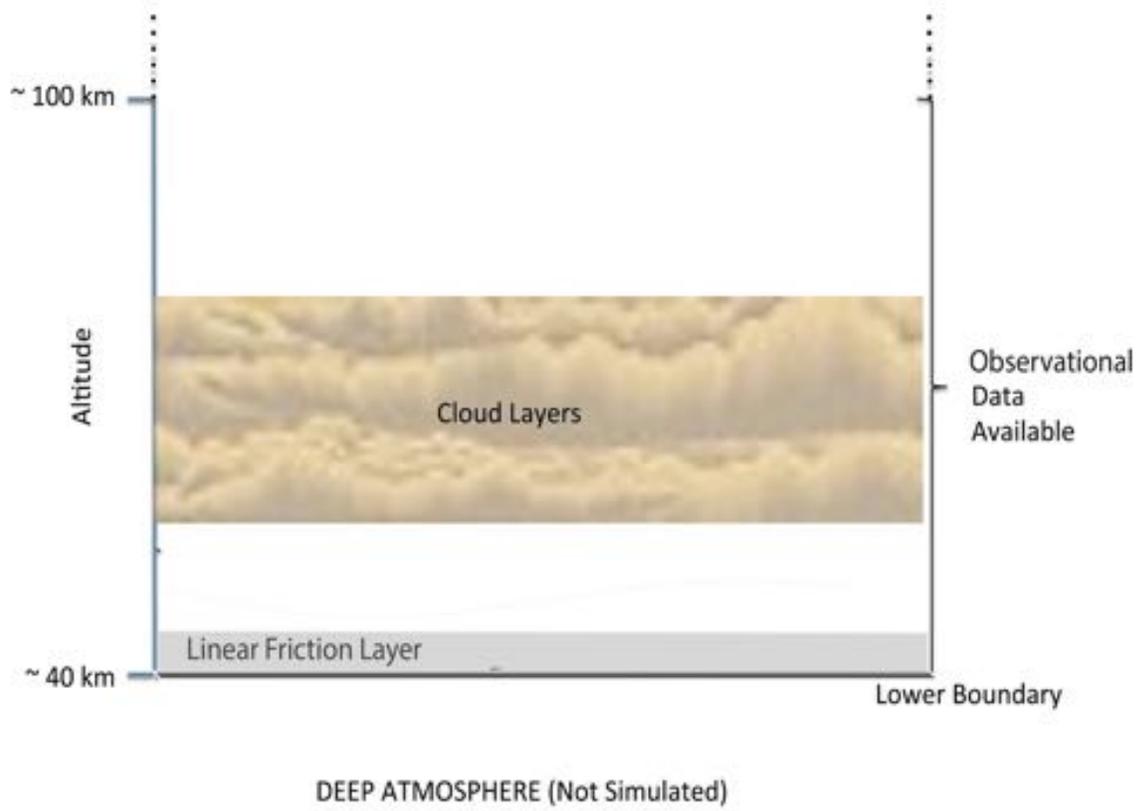

**Figure 1**: Schematic of VMM model domain

(a)

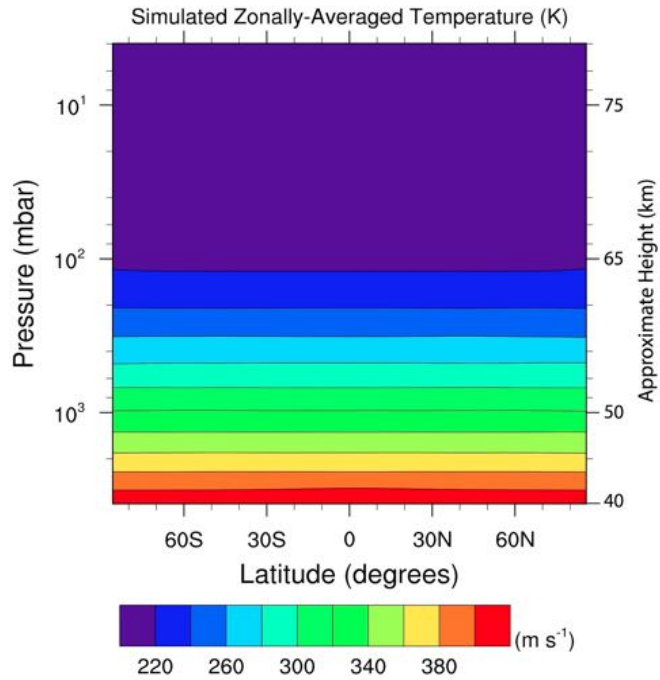

(b)

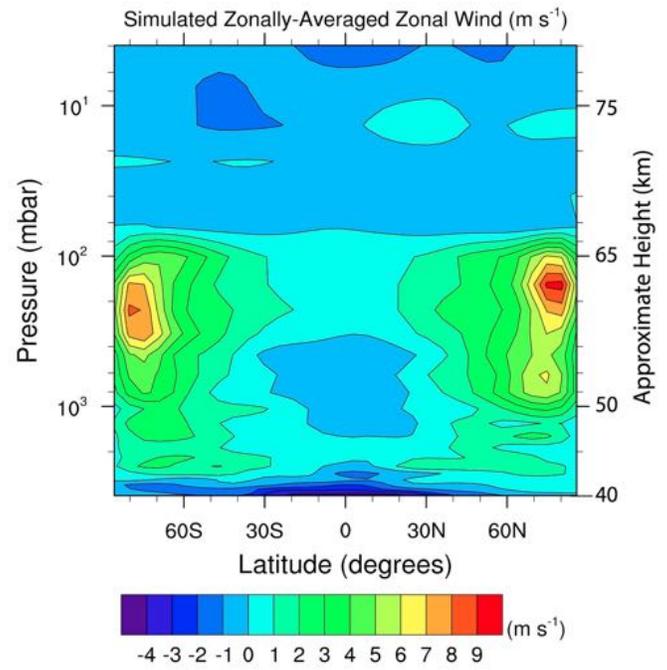

(c)

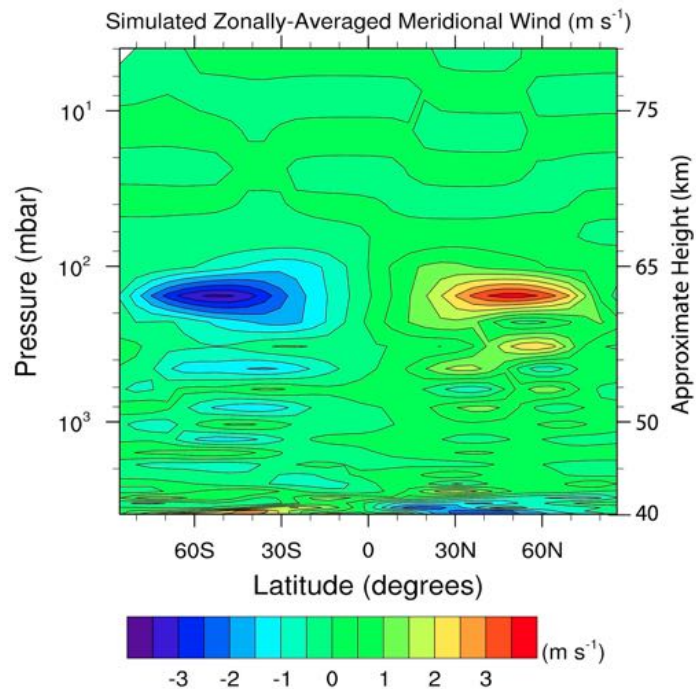

**Figure 2**: Simulated a. zonally-averaged temperature b. zonally-averaged zonal wind and
c. zonally-averaged meridional wind, as functions of latitude, pressure and approximated height,
for the baseline simulations.

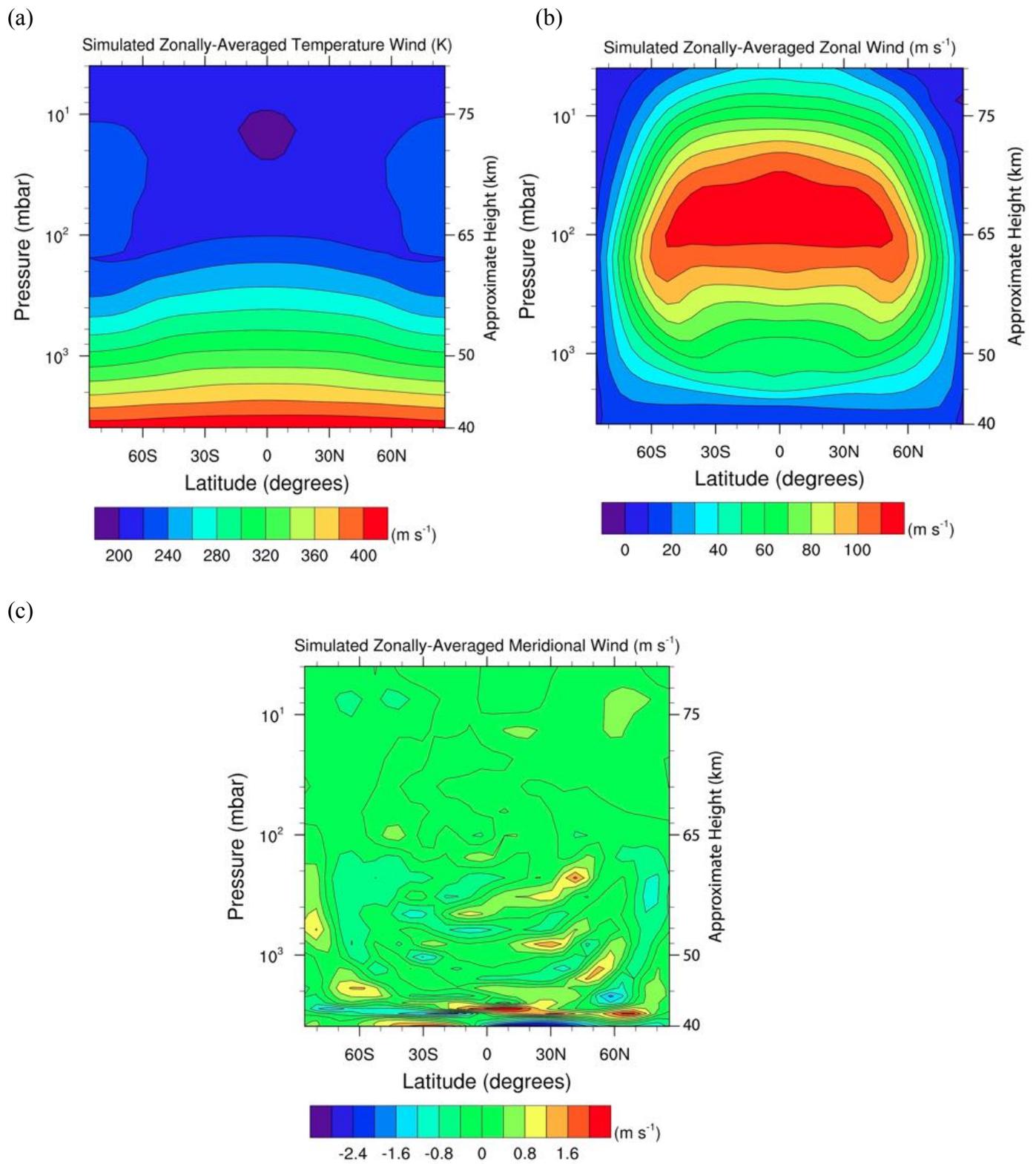

(a) Simulated Zonally-Averaged Temperature Wind (K)

(b) Simulated Zonally-Averaged Zonal Wind (m s⁻¹)

(c) Simulated Zonally-Averaged Meridional Wind (m s⁻¹)

**Figure 3**: Same as Figure 2, but for simulations which include lower boundary drag.

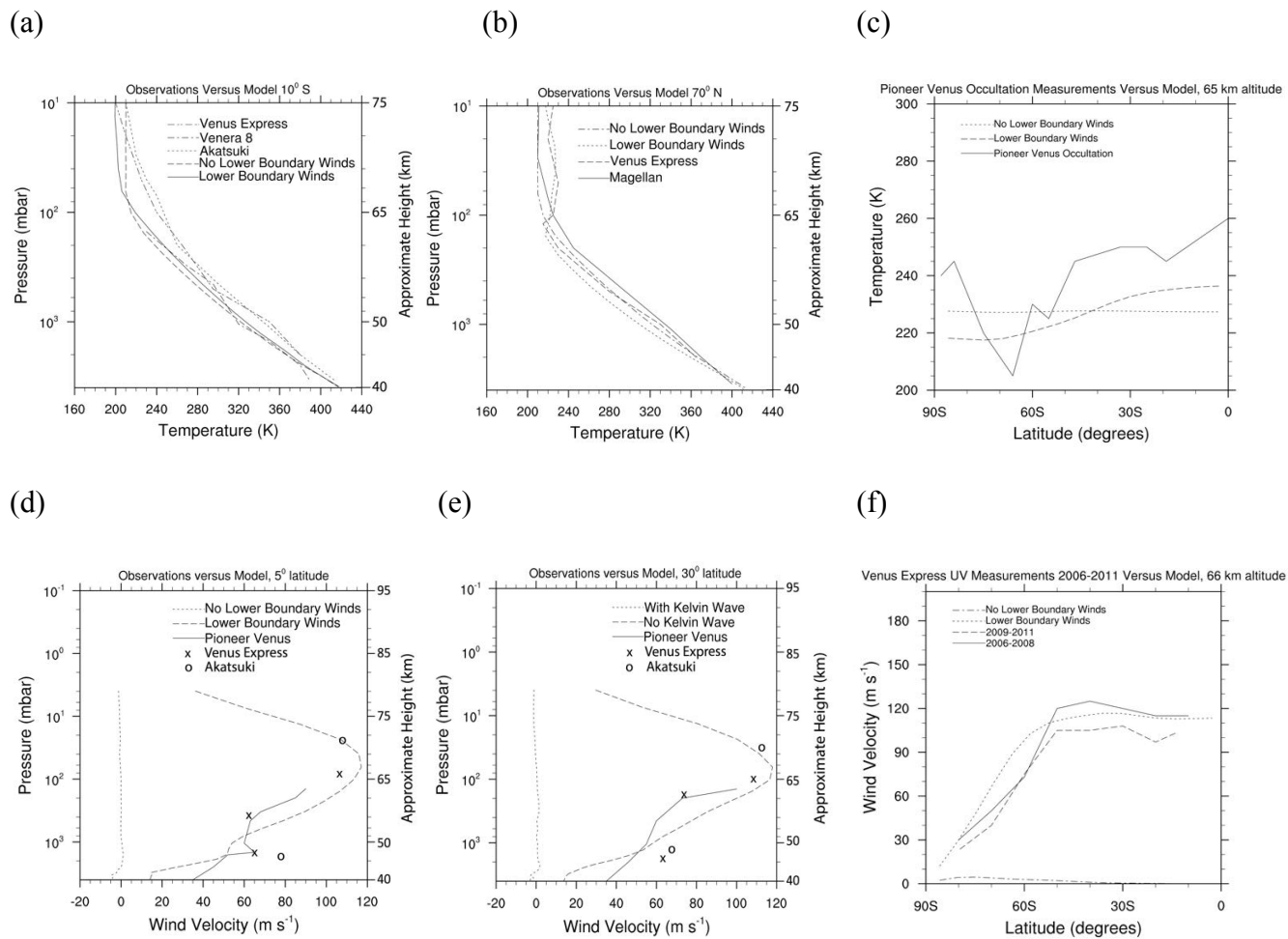

**Figure 4**: Simulated vertical profile of temperature with and without lower boundary drag compared with observations a. at $10^0$ latitude b. at $70^0$ latitude. c. Simulated latitudinal profile of temperature with and without lower boundary drag compared with observations. d. Simulated vertical profile of zonal winds with and without lower boundary drag compared with observations at $5^0$ latitude e. same as d. at $30^0$ latitude. e. Simulated latitudinal profile of zonal wind with and without lower boundary drag compared with observations.

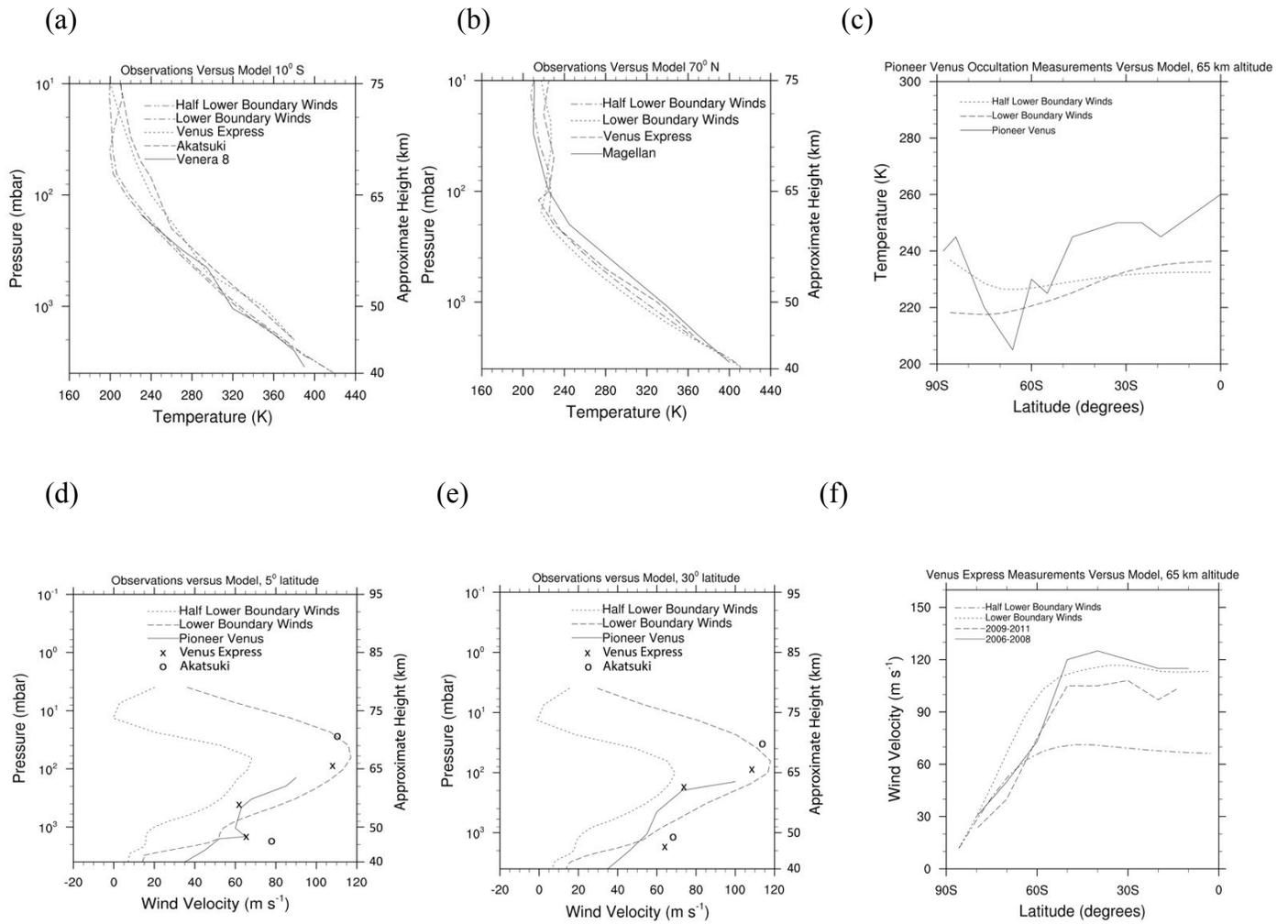

**Figure 5**: Same as Figure 4, but for simulations with original lower boundary winds and with lower boundary winds reduced to half of their original magnitudes.

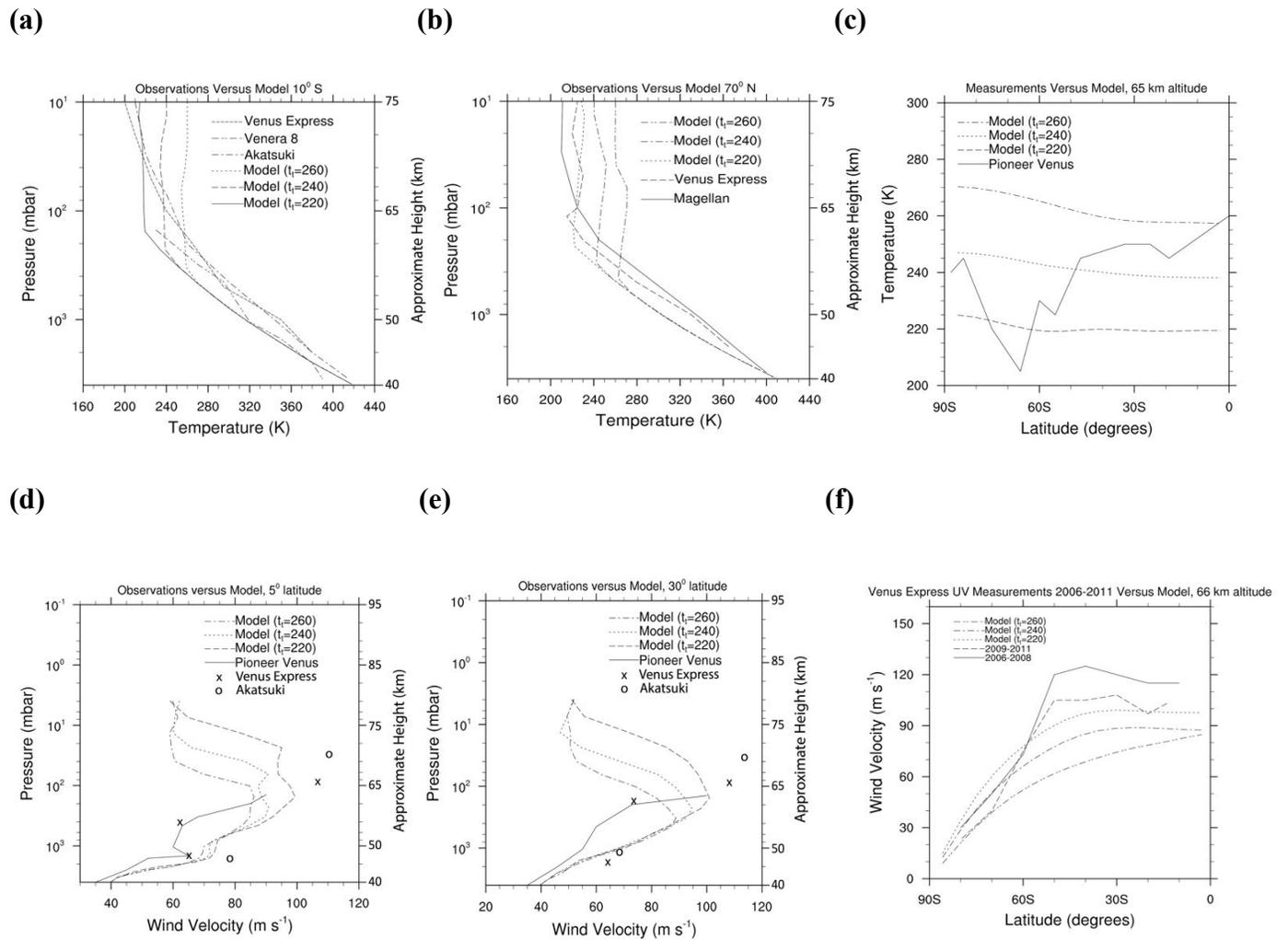

**Figure 6**: Same as Figure 4, but for simulations with tropopause temperatures of 220K, 240K and 260K respectively.

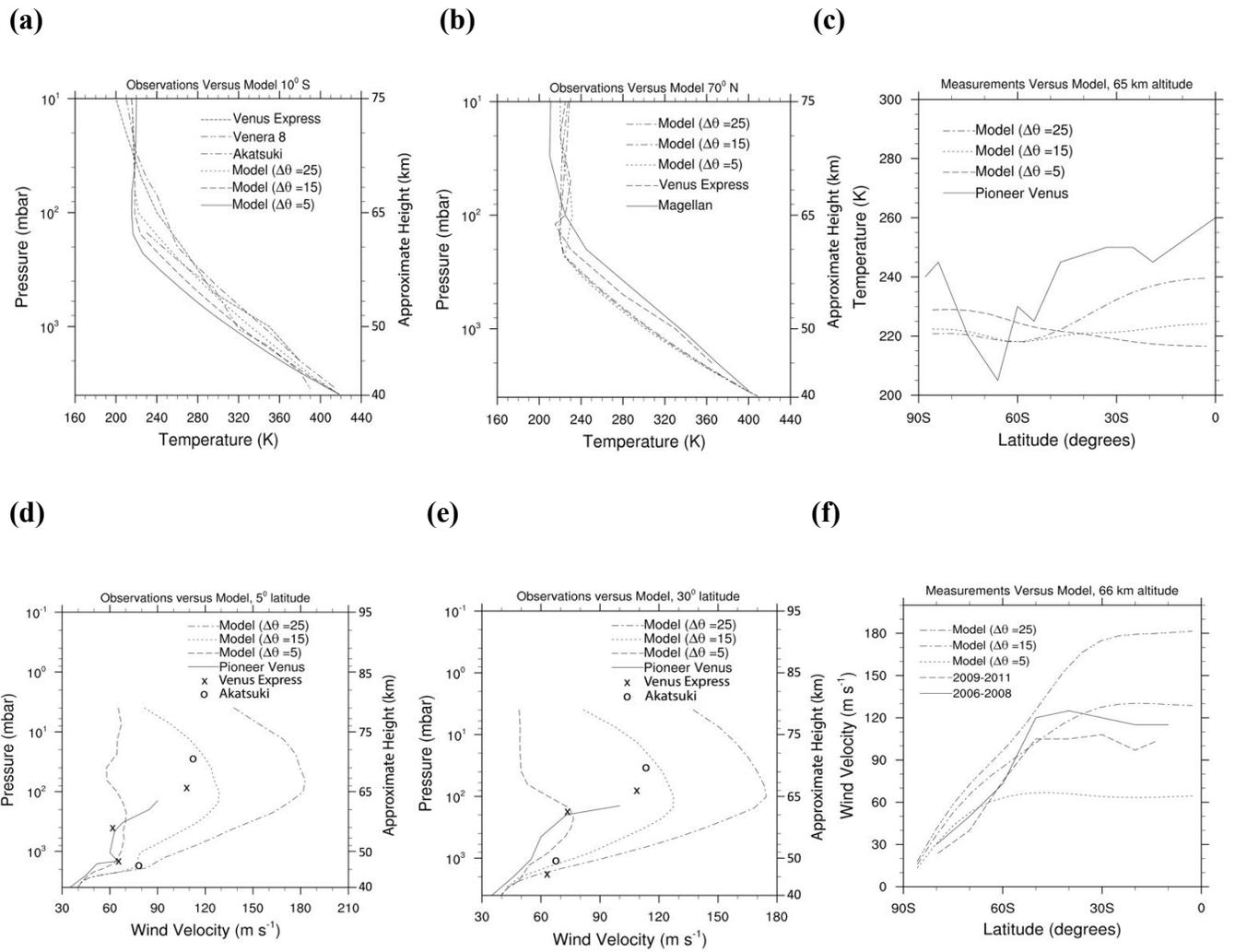

**Figure 7**: Same as Figure 4, but for simulations with a change of potential temperature with height of 5K, 15K and 25K respectively.

## Acknowledgements

H.F. Parish and J.L. Mitchell acknowledge support for this project from the National Science Foundation under award number 1614762.